\documentclass[%
 reprint,
 amsmath,amssymb,
 aps,
]{revtex4-1}

\usepackage{graphicx}
\usepackage{dcolumn}
\usepackage{bm}
\usepackage{amsmath}
\usepackage{amssymb}
\usepackage{comment}

\begin{document}

\preprint{APS/123-QED}

\title{The Information Metric on the moduli space of instantons with global symmetries}

\author{Emanuel Malek}\email{emanuel.malek@uct.ac.za}
\author{Jeff Murugan}\email{jeff.murugan@uct.ac.za}
\author{Jonathan P. Shock}\email{jonathan.shock@uct.ac.za}
 \affiliation{The Laboratory for Quantum Gravity \& Strings,\\
Astrophysics, Cosmology \& Gravity Center, \\
Department of Mathematics and Applied Mathematics,\\
University of Cape Town,\\
Private Bag, Rondebosch, 7700, South Africa\\}

\affiliation{National Institute for Theoretical Physics,\\
Private Bag X1, Matieland, 7602, South Africa\\ 
}%

\date{\today}

\begin{abstract}
In this note we revisit Hitchin's prescription \cite{Hitchin} of the Fisher metric as a natural measure on the moduli space of instantons that encodes the space-time symmetries of a classical field theory. Motivated by the idea of the moduli space of supersymmetric instantons as an emergent space in the sense of the gauge/gravity duality, we extend the prescription to encode also global symmetries of the underlying theory. We exemplify our construction with the instanton solution of the $\mathbb{CP}^{N}$ sigma model on $\mathbb{R}^{2}$.

\end{abstract}

\maketitle

\section{Introduction}
The study of the moduli space of (anti-)selfdual instanton solutions of 4-dimensional $SU(2)$ Yang-Mills theory has been an extremely rich and fruitful one, yielding a spectrum of results from Donaldson's theory of geometric invariants \cite{Donaldson:1984tm} to the recent developments in on-shell methods for scattering amplitudes (see, for example, \cite{Elvang:2013cua} and references therein). Instanton calculations also underlie key results of the AdS/CFT correspondence such as \cite{Bianchi}-\cite{Dorey02} where it was shown that $\mathcal{N}=4$ Super Yang-Mills (SYM) instantons know about the dual geometry of $AdS_5\times S^5$ from a matching of the instanton contribution of 16 fermion scatterings in the gauge theory and that of $D(-1)$ instantons in the dual geometry. It was shown that the large $N$ limit of the $k$-instanton moduli space collapses to just a single copy of $AdS_5\times S^5$ for all $k$. This very powerful matching gives us hope that instantons are a good exploratory tool of dual geometries and our aim here is to develop a generative method of finding that dual geometry using instanton moduli spaces.

The instanton moduli space generally inherits some geometric properties of the instantons that it parameterises. For example, using the $L^2$ metric, the instanton moduli space of a Hyperk\"ahler 4-manifold will be Hyperk\"ahler itself. However, for $SU(2)$ Yang-Mills the $L^2$ metric is not conformally invariant but instead relies heavily on the metric chosen for the Yang-Mills theory, not its conformal structure. Given the important role of the symmetries of gauge theories, especially conformal invariance, in the holographic context, this is a not insignificant drawback of using the $L^2$ metric when studying AdS/CFT.

In \cite{Hitchin} Hitchin proposed the Fisher information metric (see section \ref{sec.FH} for an overview of this topic) as an alternative metric to circumvent this shortcoming, albeit that this work pre-dates the AdS/CFT correspondence. In addition to being a conformally invariant metric, this metric has the desirable properties that it is complete and encodes all the spacetime symmetries of the field theory. 

In \cite{Blau} it was further suggested that using the Fisher-Hitchin metric, instantons of a CFT are a natural probe of the dynamics of the gravitational bulk dual within the AdS/CFT correspondence. Indeed, there it was demonstrated that not only was the Fisher-Hitchin metric on the 1-instanton moduli space of 4-dimensional $SU(2)$ Yang-Mills theory (Euclidean) $AdS_{5}$, but also that small perturbations of the CFT metric lead directly to linearised Einstein equations of the Fisher-Hitchin metric. Since then, information geometry has been used to define the metric on instanton moduli spaces in a number of different contexts, including at finite temperature \cite{Rey}, for $U(N)$ instantons \cite{Britto},  for $\mathbb{CP}^{1}$ and $\mathbb{CP}^{2}$ instantons \cite {Yah} and in a non-commutative context \cite{parv}. Despite this progress, it has so far not been clear how to encode the compact space of the bulk theory using the Fisher-Hitchin metric.

This article arose from our attempts to understand how to use information geometric methods to extract the full $AdS_{5}\times S^{5}$ geometry from the moduli space of instantons in $\mathcal{N}=4$ supersymmetric Yang-Mills theory. Since information about the $S^{5}$ is expected to be related to the $R$-symmetry of the SYM, we need to first understand how global symmetries of instantons are encoded in the Fisher metric. Toward this end, we warm up by studying instantons, not in the full 4-dimensional SYM but in a 2-dimensional proxy, the $\mathbb{CP}^{N}$ nonlinear sigma model that shares many of its features and reformulate Hitchin's prescription for the Fisher metric in terms of the $\mathbb{CP}^{N}$ coset current algebra.

By way of an outline, in the rest of this article we first present a quick summary of the Fisher metric and Hitchin's use of it as an alternative metric on the moduli space of 1-instanton solutions of $SU(2)$ Yang-Mills theory in four dimensions, followed by an application of the prescription to the 2-dimensional $\mathbb{CP}^{N}$ sigma model in section III. The crux of our argument, a new prescription for the metric on the moduli space of instantons in the $\mathbb{CP}^{N}$ sigma model is presented in section IV and we conclude in section V.

\section{The Fisher-Hitchin metric}\label{sec.FH}

Within the field of information theory, a measure of the distance between two (normalized) probability distributions $P$ and $Q$ over a random variable $x$ drawn from a value space, $M$, is given by the amount of information lost when one distribution is used to approximate the other. This so-called Kullback-Leibler divergence (or relative entropy) defined through \cite{KL1951,SIAbook}, 
\begin{equation}
D_{KL}(P||Q)\equiv\int_{M}\!\!\! dx\,\, P(x)\,\log\left(\frac{P(x)}{Q(x)}\right) \,,
\end{equation}
is however only a pseudo-distance as it is not symmetric under interchange of $P$ and $Q$. With this in mind, an actual distance measure on the space of parameters $\{\theta^i\}$ that describe the two distributions can be defined by taking the infinitesimal limit where $P$ approaches $Q$ so that, to leading order in $\delta \theta$,
\begin{equation}
D_{KL}(P(x;\theta)||P(x;\theta+\delta \theta))=\frac{1}{2}\delta\theta^i\delta\theta^j g_{ij}(\theta) + \ldots \,.
\end{equation}
The coefficient, 
\begin{eqnarray}
  g_{ij} \equiv \int_{M}\!\!\! dx\,\, P(x;\theta)\frac{\partial \log P(x;\theta)}{\partial\theta^{i}}
  \frac{\partial \log P(x;\theta)}{\partial\theta^{j}},
\end{eqnarray} 
defines a metric, known as the Fisher metric on the parameter space of a given family of probability density functions $P(x;\theta)$. As a canonical and illustrative example, the family of Gaussian distributions
 \begin{equation}
  P(x;\theta)=\frac 1 {\sigma\sqrt{2\pi}} e^{-\frac 1 2 \left(\frac{x-\mu}\sigma\right)^2} \,,
\end{equation}
with $\theta=(\mu,\sigma)$, yields the Fisher line element $ds^2=\left(d \mu^2+2d\sigma^2\right)/\sigma^2$. The underlying intuition is that a large standard deviation will lead to a larger overlap in two distributions, requiring more measurements to disambiguate between which of two distributions corresponds to a given source being measured. 

Hitchin's proposal was to use the Fisher metric on the instanton moduli space of field theories where the Lagrangian density evaluated on instanton solutions is used as the probability density function. The instanton moduli are then the parameters for this family of probability density functions.

Turning now to 4-dimensional SU(2) Yang-Mills theory, the conformally invariant and gauge-invariant Yang-Mills density $|F|^{2}*\!\!1 = -\mathrm{tr}\left(F\wedge *F\right)$ integrated over the 4-dimensional space, $M$, is proportional to the topological charge of the (self-dual) instanton so that 
\begin{eqnarray}
  P \equiv \frac{1}{8\pi^{2}k}|F|^{2}*\!\!1\,,
\end{eqnarray}
defines a family of normalised probability distributions on $M$ that furnishes a conformally invariant and {\it complete} metric on the instanton moduli space. The fact that the probability density function is gauge invariant means in particular that the metric is necessarily degenerate along directions in the moduli space that correspond to global gauge rotations.

\section{The $\mathbb{CP}^N$ sigma model and its instanton moduli space}
To see this construction in action in a simpler setting, let's now consider the 2-dimensional $\mathbb{CP}^N$ nonlinear sigma model 
defined through the action

\begin{equation}
 S = \int_{\mathbb{R}^{2}}\!\! d^2x\,\, \left(D_{\mu} \phi\right)^\dagger \left(D_{\mu} \phi\right) \,,
\end{equation}
where $\phi$ is an $\left(N+1\right)$-dimensional complex vector, $D_\mu \phi = \partial_\mu \phi - A_\mu \phi$ is a $U(1)$ gauge covariant derivative and $A_{\mu}$ is an auxiliary gauge connection whose equation of motion allows for it to be expressed completely in terms of the $\mathbb{CP}^{N}$ vector as $A_\mu = \phi^\dagger\partial_\mu \phi$. Completing the square in the usual way, the action can be written as
\begin{eqnarray}
  S = \mp 2\pi k + \int_{\mathbb{R}^{2}}\!\! d^{2}x\,\frac{1}{2}
  \left|\left(D_{\mu}\pm i\epsilon_{\mu\nu}D_{\nu}\right)\phi\right|^{2}
  \geq |2\pi k|\,,\nonumber
\end{eqnarray} 
with topological charge $k = \frac{i}{2\pi}\int\!\! d^{2}x\,\, \epsilon_{\mu\nu}\left(D_{\mu}\phi\right)^{\dagger}D_{\nu}\phi$. This Bogomolnyi bound saturates when $D_{\mu}\phi = \pm i\epsilon_{\mu\nu}D_{\nu}\phi$ or equivalently, with the introduction of complex coordinates $x_\pm = x_1 \pm i x_2$ when $D_{\mp} \phi = 0$. Solutions to this equation are localised, and finite action {\it i.e.} (anti-)instantons of the $\mathbb{CP}^{N}$ sigma model. A general solution takes the form
\begin{equation}
 \phi(x) = e^{i\theta} \frac{ f(x_+)}{\sqrt{f^\dagger(x_+) \cdot f(x_+)}} \,, \label{eq:kinst}
\end{equation}
and we see that (anti-)instanton solutions are expressed in terms of a (anti-)holomorphic function $f(x_\pm)$. The (anti-)instanton charge is given by the degree of the zero of $f(x_\pm)$.\\

To implement Hitchin's construction of the moduli space metric, we first need to know how many parameters parameterise an instanton for a given value of $k$. For example, the $k=1$ instanton takes the general form
\begin{equation}
 \phi = e^{i\theta} \frac{ (x_+ - a) u + \lambda v }{\sqrt{|x_+ - a|^2 + \lambda^2}} \,, \label{eq:1inst}
\end{equation}
with $a$ labelling the position of the instanton in the Euclidean $(x_+,x_-)$-plane, $\lambda$ its size and $u$, $v$ satisfying
$u^\dagger u = v^\dagger v = 1$ and $u^\dagger v = 0$ and giving the orientations of the instanton in the $\left(N+1\right)$-dimensional complex vector space.
Asymptotically, for the instanton to be a finite action solution, $A_\mu$ must be pure gauge and $\phi$ constant up to the $U(1)$ gauge symmetry. In other words the asymptotic data for the instanton is that
\begin{equation}
 \begin{split}
 \lim_{|x|\rightarrow \infty} \phi &= \phi_0 e^{i \theta} \,, \\
 \lim_{|x|\rightarrow \infty} A_\mu &= i \partial_\mu \theta \,.
 \end{split}
\end{equation}
Zero modes define nearby solutions with the same asymptotic data. Since the $U(1)$ gauge field $A_\mu$ is higgsed, the global part of the $U(1)$ is not a zero mode. Similarly, the global orientation of $\phi$ in the $\left(N+1\right)$-dimensional complex vector space is fixed by the asymptotic data and hence $u$ is not a zero mode leaving the remaining zero modes as the complex position $a$, the real scale $\lambda$ and the orientation vector $v$. The constraints $v^\dagger v = 1$ and $u^\dagger v = 0$ imply $2N - 1$ independent components for $v$ and therefore a total of $2(N+1)$ zero modes in $\mathbb{CP}^{N}$.

Plugging the 1-instanton solution back into the the Lagrangian density gives the on-shell expression
\begin{equation}
{\mathcal L}= \frac{\lambda^2}{\left(\lambda^2 + |x_+-a|^2\right)^2} \,.
\end{equation}
Following Hitchin's prescription, the normalised Lagrangian density, $\widetilde{\mathcal{L}}$, defined so that $\int\! d^2x\,\widetilde{\mathcal L}=1$ is the probability density function. The associated Fisher-Hitchin metric
\begin{equation}
g_{ab}=\int \!d^2x\,\, \widetilde{\mathcal L}\,\left(\partial_a\log\, \widetilde{\mathcal L}\right)\left(\partial_b\log \widetilde{\mathcal L}\right)\,,
\end{equation}
produces the line element
\begin{equation}
ds^2=4\frac{da_1^2+da_2^2+d\lambda^2}{3\lambda^2}\,.
\end{equation}
This is an immediate consequence of the $SO(1,3)$ conformal symmetry of the planar $\mathbb{CP}^N$ sigma model. This symmetry acts on $a$ and $\lambda$ which are the only zero modes appearing in the probability density $\widetilde{\mathcal L}$. The $v$ zero modes have been ``traced out'' and give degenerate directions in the Fisher metric.
Consequently, while the conformal nature of the underlying classical field theory has been captured, since $\widetilde{{\mathcal L}}$ is a singlet under all global symmetries, we have no chance of capturing the internal symmetries of the instantons in the Fisher-Hitchin metric. The question then arises as to how we can possibly encode the global symmetries of the instanton in the Information metric? This is particularly important if one considers the instanton moduli space of $\mathcal{N}=4$ Super Yang-Mills where we would  hope to capture both the AdS$_5$ as well as the $S^5$ parts of the holographic dual. 

\section{An Information metric as a functional of the $\mathbb{CP}^N$ coset current}
We now discuss a way for the Fisher metric to retain information about the internal symmetries. As is clear, we must use a density function that is charged under those internal symmetries. Such a density function cannot yield a probability density function. It does, however, behave like an amplitude that ``squares'' to the probability density. There are a couple of different such amplitudes that come to mind. For example, one may use $D_+ \phi$ evaluated on $k=1$ instantons. However, since we want our prescription to be as broad as possible, we will exploit the fact that the $\mathbb{CP}^{N}$ sigma model can be formulated as a coset sigma model on the coset 
\begin{equation}
 \mathbb{CP}^N=\frac{U(N+1)}{U(N)\times U(1)} \,,
\end{equation}
and utilise the corresponding coset current in our proposed modification of the Fisher metric. In this way, our construction is expected to generalise to any coset model. For concreteness, we will make our argument with the 1-instanton of the $\mathbb{CP}^{N}$ model where we define an Information metric in terms of the coset current $J$ evaluated on the $k=1$ instanton as
\begin{equation}\label{eq.newig}
 G_{ab} \equiv 4 \int_{\mathbb{R}^{2}}\!\!\! d^2x\,\, \mathrm{tr}\left[ \partial_{(a} \hat{J}^\dagger\,\partial_{b)} \hat{J}\right] \,,
\end{equation}
with $\hat{J}$ a normalised current associated to $J$ such that $\int d^2x\,\, \mathrm{tr}\left(\hat{J}^\dagger \hat{J}\right) = 1$. 

Although we are not aware of such a definition in existing Information Theory literature, we can motivate the above definition by noting that (i) the coset current $J$ does indeed behave like an amplitude by virtue of its relation to the Lagrangian as $\mathcal{L} = \mathrm{tr} \left(J^\dagger J\right)$, and (ii)
when the naive replacement $J\sim\sqrt{\mathcal{L}}$ is placed into (\ref{eq.newig}) we get back the normal definition of the Fisher metric. As we demonstrate below however, this new formulation also allows for internal moduli to be encoded when they are present. That this definition has the same overall power of $J$ as the definition in \cite{Hitchin} is crucial for the integral to be well defined.

Recall that to construct the coset current for a general coset space $G/H$, we start with the Maurer-Cartan form of $G$ given as $g^{-1} dg$, where $g \in G$ is a general group element. The Maurer-Cartan form itself is an element of $\mathfrak{g}$, the Lie algebra of $G$. An $H$-invariant current is then obtained by projecting onto $\mathfrak{m}$, the orthogonal complement to the Lie algebra $\mathfrak{h}$ of $H$, with respect to the Killing form on $\mathfrak{g}$. We denote this projection by
\begin{equation}
 J = g^{-1} dg|_{\mathfrak{m}} \,.
\end{equation}

For the purpose of calculating the coset current, we can use any coset representative. For $\mathbb{CP}^{N}$ we will use
\begin{equation}
g =  \begin{pmatrix}
  e^{-i\theta} \left( \delta^\alpha{}_\beta + b^\alpha b^*_\beta \left(\sin\rho - 1\right)\right) & b^\alpha \cos\rho \\
  - b^*_{\beta} \cos\rho & e^{i\theta} \sin\rho
 \end{pmatrix} \,,
\end{equation}
for some $\rho$, $\theta \in \mathbb{R}$ and a complex vector $b^\alpha$ satisfying $b^\alpha b_\alpha^* = 1$, where the indices $\alpha, \beta = 1, \ldots N$ are $U(N)$ indices. We now need to evaluate $g$ and $J$ on instanton solutions. To do this, let us first relate the matrix $g$ to the $\mathbb{CP}^N$ vector $\phi$ as
\begin{equation}
\phi = g\begin{pmatrix}
 0 \\ 1
 \end{pmatrix} =\begin{pmatrix}
  b^\alpha \cos \rho \\
  e^{i\theta} \sin \rho
 \end{pmatrix} \,.
\end{equation}
We now equate this with the instanton solution \eqref{eq:1inst} to find
\begin{equation}
  \begin{pmatrix}
    b^\alpha \cos \rho \\
    e^{i\theta} \sin \rho
   \end{pmatrix}=\frac{1}{\sqrt{r^2+\lambda^2}}\left(\lambda v+y_+u\right) \,,
\end{equation}
where we have introduced the variables $y_+ = x_+ - a$ and $r = |y_+|$. This allows us to identify
\begin{equation}
 \begin{split}
  u &= \begin{pmatrix}
   0 \\ 1
  \end{pmatrix} \,, \qquad
  v = \begin{pmatrix}
   b^\alpha \\ 0
  \end{pmatrix} \,, \\
  y_+ &= r e^{i\theta} \,, \qquad \cos \rho = \frac{\lambda}{\sqrt{r^2+\lambda^2}} \,,
 \end{split}
\end{equation}
where $b^\alpha$ is constant for the 1-instanton solution. We can now calculate the coset current, $J$, evaluted on the instanton solutions. Firstly, we find
\begin{equation}
 g^{-1} dg = \begin{pmatrix}
  - id\theta \left( \delta^\alpha{}_\beta + b^\alpha b^*_\beta \cos^2\!\rho \right) & - q b^\alpha \\
  q^* b^*_\beta & i d\theta \sin^2\!\rho
  \end{pmatrix} \,, \label{eq:MCForm}
\end{equation}
where we have defined
\begin{equation}
 q = e^{i\theta} \left( d\rho + \frac{i}{2} \sin 2\rho~ d\theta \right) = \frac{\lambda}{r^2 + \lambda^2} dx_+ \,.
\end{equation}
Projecting \eqref{eq:MCForm} onto $\mathfrak{m}$ gives the current
\begin{equation}
 J = g^{-1} dg|_{\mathfrak{m}} = \begin{pmatrix}
  0 & - q b^\alpha \\
  q^* b_\beta^* & 0
 \end{pmatrix} \,.
\end{equation}

Finally, substituting into our prescription for the Information metric gives
\begin{equation}
 \begin{split}
  G_{ab} 
  &= \frac{4}{\pi} \int d^2x \partial_{(a} \left( q_\mu b^\alpha \right) \partial_{b)} \left( q_\mu^* b^*_{\alpha} \right) \,.
 \end{split}
\end{equation}
We can parameterise the $N$-dimensional complex unit-vector $b^\alpha$ recursively in terms of a family of $n$-dimensional unit-vectors, $\mathbf{c}_{n}$:
\begin{equation}
 b^\alpha = \left(\mathbf{c_N}\right)^\alpha \,, \qquad \mathbf{c}_n = \begin{pmatrix}
  e^{i\rho_n} \cos\theta_{n-1} \\ \mathbf{c}_{n-1} \sin \theta_{n-1}
 \end{pmatrix} \,, \qquad c_1 = e^{i\rho_1} \,.
\end{equation}
The resultant metric is $AdS_3 \times M^{2N-1}$. The moduli $a_1$, $a_2$, $\lambda$ are coordinates on the same $AdS_3$ metric as for the Fisher-Hitchin prescription, whereas the moduli $\rho_1$, \ldots, $\rho_N$, $\theta_1$, \ldots, $\theta_{N-1}$ parameterise a $(2n-1)-$dimensional compact space with metric:
\begin{equation}
 ds^2 = \frac{4}{3\lambda^2} \left(da_1^2 + da_2^2 + d\lambda^2 \right) + 4 d\Omega_N^2 \,,
\end{equation}
where $d\Omega_N^2$ is defined recursively in terms of
\begin{equation}
 d\Omega_n^2 = d\theta_{n-1}^2 + \cos^2\theta_{n-1} d\rho_{n}^2 + \sin^2\theta_{n-1} d\Omega_{n-1}^2 \,,
\end{equation}
and $d\Omega_1^2 = d\rho_1^2$. For $N=1$ and $N=2$, the compact space is nothing but an $S^1$ and $S^{3}$ in Hopf coordinates.

\section{Discussion}
We have shown here that Hitchin's proposal for a gauge and conformally-invariant, geodesically complete metric on the moduli space of instantons, while superior to the usual $L^{2}$ metric in these respects, is nevertheless insufficient to capture information about the internal symmetries of the gauge theory. In a simple extension of the information metric, we have argued that a new current formulation of the metric on the space of instanton solutions does indeed capture both isometric information as well as information about the global internal symmetries. As a proof of principle, in this article, we have shown that the $1$-instanton moduli space of the planar $\mathbb{CP}^N$ nonlinear sigma model is equipped with an $AdS_3\times M^{2N-1}$ metric.\\

As discussed in the introduction, our motivation for this study comes from trying to understand to what extent gauge theory instantons are a good probe of dual geometry and, more specifically, how the instanton moduli space is related to the emergent dual spacetime. In this light, there are a number of further avenues to persue. Foremost among these is an understanding of the moduli space of supersymmetric instantons, both in nonlinear sigma models as well as gauge theory. Understanding how the R-symmetry of the latter is coded in the information metric on the moduli space is, of course, key to our end goal.\\

Ideas from information theory, in particular the geometry of information manifolds are currently enjoying a resurgence in high energy physics, due in no small part to the rapid development of the field of quantum entanglement and entanglement entropy using holographic methods \cite{Nozaki:2012zj, Lin:2014hva,Miyaji:2015mia}. It would be most interesting to understand how these two, seemingly disparate, threads in which information geometry plays such a key role in the emergence of spacetime, may be related to each other.

\section{Acknowledgements}
We are indebted to Michael Abbott for useful discussions regarding the coset construction in particular. In addition we are grateful for insight from Tslil Clingman, Johanna Erdmenger, Hiroaki Matsueda, Andrew Singleton, Koenraad Schalm and David Tong. JS is grateful to the KITPC and IHEP in Beijing who hosted him while this project was nearing completion. This work is based on the research supported in part by the National Research Foundation of South Africa (Grant Numbers 87667, and 90519). Opinions, findings and conclusions or recommendations expressed in any publication generated by the NRF supported research is that of the author(s), and that the NRF accepts no liability whatsoever in this regard.

\end{document}